\documentclass[aps,prl, showpacs,twocolumn]{revtex4}
\usepackage{amssymb}
\usepackage{amsmath}
\usepackage{graphicx}
\usepackage{epsfig}

\begin{document}

\title{Engineering antiferromagnetic Heisenberg spin chains for maximizing
of the groundstate entanglement}
\author{ R. Xin $^{1}$, Z. Song$^{1,a}$ and C. P. Sun$^{1,2,a,b}$ }
\affiliation{$^{1}$Department of Physics, Nankai University, Tianjin 300071, China}
\affiliation{$^{2}$ Institute of Theoretical Physics, The Chinese Academy of Science,
Beijing, 100080, China}

\begin{abstract}
We study the correlation function and concurrence for the
eigenstates with zero spin of engineered Heisenberg models to
explore the entanglement property. It is shown that the total
nearest neighbor (NN) correlation function of zero-spin
eigenstates reaches its local extremum when the coupling strength
is uniform, and correspondingly the groundstate entanglement of
$d$-D cubic AF Heisenberg model is locally maximized. Moreover,
numerical calculations for a $N$-site quantum spin ring with
cosinusoidally modulated exchange coupling, i.e.
$J_i=J(1+\cos(2\pi i/N))$, indicate that the uniform coupling is
not the unique optimal distribution for maximizing the groundstate
entanglement and this modulation of interactions can induce the
longer range entanglement.
\end{abstract}

\maketitle

\emph{Introduction.} Studies of various quantum spin models are very
important in understanding the properties of realistic solid state systems.
Numerical and analytical studies have shown that the groundstate properties
are sensitive to the geometry of the lattice ~\cite{1,2,3,4,5} and the
distribution of the exchange couplings between two neighbor spins \cite%
{6,Chris,song1,song2}. Although much effort has already been made
to investigate the properties of the ground state, the exact
result about it is still rare. For instance, a spin-$1/2$
Heisenberg ring is one of the nontrivial, relatively simple
physical systems. It was the first one dimensional quantum model
solved exactly by a straightforward diagonalization of the
Hamiltonian and has been used as a testing ground for many
theoretical approaches. Bethe~ found the eigenvalues and
eigenfunctions of the corresponding Hamiltonian \cite{7}, but the
obtained exact eigenfunctions are so complicated that they are
almost useless for calculating some physical quantities, such as
something relevant to correlation functions, to reveal all the
characteristic properties of the ground state. Recently, quantum
entanglement was proposed as a new type of physical resource,
which is desired to depict the features of the ground state, like
the correlation function or other conservative quantities. It was
found that the ground state of an AF Heisenberg ring
possesses maximal entanglement compared to that of excited states \cite%
{Wang} and the entanglement is believed to have somthing to do
with the quantum phase transitions\cite{Lin1}.

The present paper explores the pairwise entanglement of the
zero-spin ground state of the Heisenberg model with certain
engineered inhomogeneous coupling constants. We first make a
simple observation that the total nearest neighbor (NN)
correlation function of zero-spin eigenstate (with vanishing
components of total spin) is locally maximized for the uniformly
distributed
coupling strengths, while the groundstate entanglement of the $d$%
-D cubic antiferromagnetic (AF) Heisenberg model can locally
maximize in this case, correspondingly. However, the uniform
distribution of coupling constants is not the unique optimal one
for maximizing the groundstate entanglement in the case with
engineered coupling constants. Namely, the homogeneity of
couplings is only the sufficient conditions rather than a
necessary condition in maximizing the groundstate entanglement. To
demonstrate this point of view, we study the $N$-site ring
system with cosinusoidally modulated exchange coupling, i.e., $%
J_{i}=J(1+\cos (2\pi i/N))$ in both analytical and numerical
approaches. We discovered that the ground state \ with varying
couplings $J_{i}$ and that with the fixed coupling $J$ are nearly
identical.

\emph{Maximization of the groundstate entanglement.} The Hamiltonian $%
H=\sum_{\left\langle ij\right\rangle }J_{ij}\mathbf{S}_{i}\cdot \mathbf{S}%
_{j}$ of Heisenberg spin model can be written in terms of the spin operator $%
\mathbf{S}_{i}$ at $i$th site, where $J_{ij}$ is the coupling
constant of exchange interaction, which is trivially restricted to
be \emph{nonzero} in this paper. Our studies will focus on those
states with vanishing components of total spin, or called the zero
spin states. In this sense the relationship between correlation
and concurrence has been well established \cite{Wootters}. To
describe the behavior of correlation one can define the
nearest neighbor (NN) correlation function%
\begin{equation}
F_{0}(J_{ij})=\frac{1}{N}\sum_{\left\langle ij\right\rangle }\langle \mathbf{%
S}_{i}\cdot \mathbf{S}_{j}\rangle _{0},
\end{equation}%
for the eigenstate $|\psi _{0}\rangle $ with zero spin. The sandwich $%
\langle \mathbf{S}_{i}\cdot \mathbf{S}_{j}\rangle _{0}=\langle \psi _{0}|%
\mathbf{S}_{i}\cdot \mathbf{S}_{j}|\psi _{0}\rangle $ is defined as the
expectation value of $\mathbf{S}_{i}\cdot \mathbf{S}_{j}$ in the zero-spin
eigenstate $|\psi _{0}\rangle $. Formally, it is proportional to the average
of the Hamiltonian by assuming the identical coupling constants. What we
concern is the $J_{ij}$-dependent behavior of the correlation function,
which can be characterized by the \textit{extremum of }$F_{0}(J_{ij})$.

Now, for a zero-spin eigenstate, we show that $F_{0}(J_{ij})$ can
reach its extremum when all the exchange constants are identical,
$J_{ij}=J$, if $F_{0}(J_{ij})$ is analytical at this point.
Actually, the eigen energy can be written as
\begin{equation}
E_{0}=\langle H\rangle _{0}=\sum\limits_{\left\langle ij\right\rangle
}J_{ij}\langle \mathbf{S}_{i}\cdot \mathbf{S}_{j}\rangle _{0}.
\end{equation}%
Differentiating the above equation with respect to an arbitrary coupling
strength $J_{kl}$, one can get $\partial E_{0}/\partial
J_{kl}=\sum_{\left\langle ij\right\rangle }J_{ij}\partial \langle \mathbf{S}%
_{i}\cdot \mathbf{S}_{j}\rangle _{0}/\partial J_{kl}+\langle \mathbf{S}%
_{k}\cdot \mathbf{S}_{l}\rangle _{0}.$ On the other hand, together
with the Feynman--Hellman theorem
\begin{equation}
\frac{\partial E_{0}}{\partial J_{kl}}=\langle \frac{\partial H}{\partial
J_{kl}}\rangle _{0}=\langle \mathbf{S}_{k}\cdot \mathbf{S}_{l}\rangle _{0}
\end{equation}%
it indicates that $\sum_{\left\langle ij\right\rangle
}J_{ij}\partial \langle \mathbf{S}_{i}\cdot \mathbf{S}_{j}\rangle
_{0}/\partial J_{lk}=0$. Obviously, when all the coupling
strengths $J_{ij}=J$, we get $\partial F_{0}/\partial J_{ij}=0$,
which means that $F_{0}$ has an extremum for the uniform coupling
strength. Notice that this conclusion is always true no matter the
system is bipartite or non-bipartite lattice and the state is
ground state or excited state. However, $J_{ij}=J$ may not be the
unique distribution for $F_{0}$ to reach the extremum. We will
find that there may exist a periodic coupling strength
distribution in 1-D system which induces the same ground state
approximately as that with uniform coupling distribution.

Now we apply the above conclusion to the bipartite lattice with
$J_{ij}>0$ and $N_{A}=N_{B}$, where $N_{A}$, $N_{B}$ are the
numbers of the sites belonging to sublattices $A$ and $B$.
According to Lieb's theorem~\cite{lieb}, the ground state is
singlet and has $S=0$, which ensures that $F_{g}$ reaches its
minimum $F_{g}(1)=E_{g}/N$ at the point $J_{ij}=1$, where the
zero-spin state $|\psi _{0}\rangle $ is replaced by the ground state $%
|g\rangle $. Furthermore, we consider the groundstate entanglement
for the AF Heisenberg model on a $d$-D cubic lattice with
translational symmetry in all directions. Here the symmetry
specifies the \emph{geometry} of the lattice only, i.e., the
Hamiltonian may not have the translational symmetry. Since the
upper bound of $E_{g}$ is $-\sum_{\left\langle ij\right\rangle
}1/4$ in the vicinity of the point $J_{ij}=1$, we have the
inequality $F_{g}(1)<-(1/N)\sum_{\left\langle ij\right\rangle }1/4$ from $%
\langle \mathbf{S}_{i}\cdot \mathbf{S}_{j}\rangle _{g}<-1/4$. On the other
hand, the pairwise concurrence for such system is $C_{g}=(1/2)\max
\{-(4/N)\sum_{\left\langle ij\right\rangle }(\langle \mathbf{S}_{i}\cdot
\mathbf{S}_{j}\rangle _{g}-1),0\}=$ $(1/2)\max
\{-[4F_{g}(1)+(1/N)\sum_{\left\langle ij\right\rangle }1],0\}$ \cite%
{Wootters}, or
\begin{equation}
C_{g}=-\frac{1}{2}[4F_{g}(1)+\frac{M}{N}],
\end{equation}%
which maximizes in the vicinity of the point $J_{ij}=1$, where $%
M=\sum_{\left\langle ij\right\rangle }1$ is the link numbers.
Therefore we get the conclusion that the ground state has locally
maximal pairwise entanglement when the exchange interactions
distribute uniformly. A similar conclusion has been obtained for
the $XXZ$ model at the isotropic point \cite{Yang, Lin2}.

Notice that the above statement does not mean that the concurrence is the
maximum in the whole range of coupling constants, but just in the vicinity
of the uniform point. In order to illustrate this, we investigate a simple
spin model, the AF Heisenberg ring with alternative coupling constant. The
Hamiltonian of $N$-site ring is
\begin{equation}
H=\sum_{i\in odd}\mathbf{S}_{i}\cdot \mathbf{S}_{i+1}+\sum_{i\in even}J%
\mathbf{S}_{i}\cdot \mathbf{S}_{i+1}.
\end{equation}%
It is well known that the value of the $NN$ pairwise concurrence
is $0.386$ for
$J=1$ \cite{Wootters}. On the other hand, in the limit cases of $J=0$ and $%
\infty $, the exact ground states are $\phi _{1}=[12][34]\ldots
\ldots \lbrack N-1N]$, and $\phi _{2}=[23][45]\ldots \ldots
\lbrack N1]$ respectively, where $[ij]$ denotes a resonant valence
bond (RVB) of two spins located at the lattice sites $i$ and $j$
\cite{RVB}. Both $\phi _{1}$ and $\phi _{2}$ \ has a same
concurrence $0.5$, which is larger than that of the state at
uniform point. This demonstrates that the concurrence is not the
maximum in the whole range of coupling constants. The groundstate
concurrences for small systems with $N=8,10$ and $12$ are
calculated by exact diagonalization method and plotted in Fig.1.
It indicates that the concurrences takes its local maximum at the
uniform point.

\emph{Cosinusoidally modulated systems.} As mentioned in advance, the
uniform system may not be the unique optimal one possessing a maximal
groundstate entanglement. A simplest case is that there may exist a model
with non-uniform coupling strength distribution, which has the same
ground state as that of the uniform one. We now show this observation in the $%
XY$ spin ring exactly and the $XXX$ spin ring approximately.

\vspace*{-2.0cm}

\begin{figure}[h]
\hspace{24pt}\includegraphics[width=7.5cm,height=11cm]{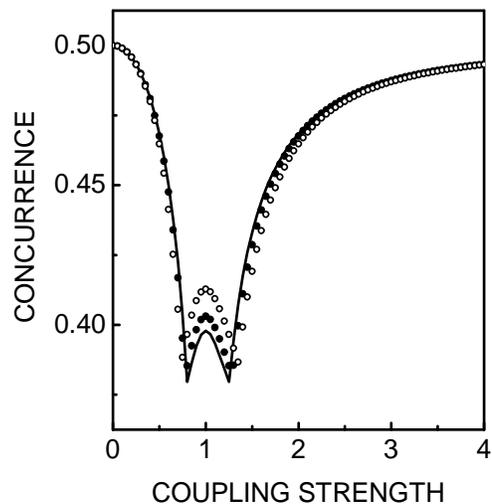}
\par
\vspace*{-1.5cm}
\caption{\textit{The groundstate concurrences vs the coupling strength for
the alternative Heisenberg ring systems with $N=8$ (circle), $10$ (solid
circle) and $12$ (solid line). It shows that the uniform coupling point is
the local maximum. }}
\end{figure}

We consider the two cases of spin-$1/2$ Heisenberg model with the
anisotropy parameter $\Delta =0,1$ on an $N$-site ring. With the
cosinusoidally modulated exchange couplings, i.e. $J_{i}=J(1+\cos (2\pi i/N))$%
, the Hamiltonian $H=H_{0}+H_{add}$ can be separated into two parts%
\begin{equation}
H_{0}=J\sum_{i}^{N}(S_{i}^{x}S_{i+1}^{x}+S_{i}^{y}S_{i+1}^{y}+\Delta
S_{i}^{z}S_{i+1}^{z}),
\end{equation}%
and
\begin{equation}
H_{add}=J\sum_{i}^{N}\cos (\frac{2\pi i}{N}%
)(S_{i}^{x}S_{i+1}^{x}+S_{i}^{y}S_{i+1}^{y}+\Delta S_{i}^{z}S_{i+1}^{z}).
\end{equation}%
where $2\pi i/N$ is a factor to determine the profile of the additional
Hamiltonian. $H$ was a pure toy model before the array of quantum dots is
considered as a media to transfer the quantum states~\cite{array,Chris,song1}%
. In Fig.2 $H$ and $H_{0}$ are illustrated schematically. We will see that $%
H_{add}$ has a subtle relation with $H_{0}$ in their ground
states.

\begin{figure}[h]
\includegraphics[bb=55 140 540 600, width=4cm,clip]{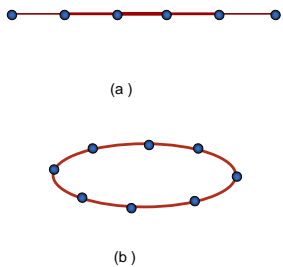}
\caption{\textit{Schematic illustration of cosinusoidally modulated (a) and
uniform (b) NN coupling spin ring systems. The two most separated sites in
(a) is the nearest neighbor in (b).}}
\end{figure}

The simplest case $\Delta =0$ \ is first considered as the so
called $XY$ model or the hardcore boson model. It is well known
that if $N/2$ is odd the spectrum of the $XY$ Heisenberg spin ring
in the subspace with $S_{z}=0$ is reduced to the model of
non-interacting spinless fermion system. The
reduced Hamiltonian consists of two parts $H_{0}^{XY}=(J/2)%
\sum_{i}^{N}(a_{i}^{\dagger }a_{i+1}+h.c)$ and $H_{add}^{XY}=(J/2)$ $%
\sum_{i}^{N}\cos (2\pi i/N)(a_{i}^{\dagger }a_{i+1}+h.c)$, written
respectively in terms of the spinless fermion operata $a_{i}^{\dagger }$ at $%
i$th site. In $k$-space, by using the discrete Fourier transformation $%
a_{l}^{\dagger }=1/\sqrt{N}\sum_{l}^{N}a_{k}^{\dagger }\exp (ikl),$( $k=2\pi
n/N, n=0,\pm 1,\pm 2,...,$ $\pm (N/2-1),$ $N/2)$, they can be re-written as%
\begin{equation*}
H_{0}^{XY}=J\sum_{k}\cos ka_{k}^{\dagger }a_{k}
\end{equation*}%
and
\begin{equation}
H_{add}^{XY}=\frac{J}{2}\sum_{k}[e^{-i\frac{\pi }{N}}\cos (k+\frac{\pi }{N}%
)a_{k}^{\dagger }a_{k+\frac{2\pi }{N}}+h.c].
\end{equation}

The physics of the above Hamiltonian is quite obvious. The additional
Hamiltonian $H_{add}^{XY}$ describes an $N$-site chain system with nearest
neighbor hopping in $k$ space. Notice that the $k$-dependent hopping
integral $\propto \cos (k+\frac{\pi }{N})$ vanishes at points $k=\pm k_{f}$,
where $k_{f}=\pi /2-\pi /N$ is the "fermi point" ( surface ) for the one
dimensional half-filled spinless fermion model $H_{0}^{XY}$. It is easy to
find that the particles are confined either in the regions $|k|\leq k_{f}$
or $|k|>k_{f}$. Actually, it is easy to prove that the particle number in
the region $|k|\leq k_{f}$ ( or $|k|>k_{f}$ ) is a good quantum number for $%
H_{add}^{XY}$, i.e., $[\ \mathbf{n},H_{add}^{XY}\ ]=0$, where $\mathbf{n}%
=\sum_{k\in \{|k|\leq k_{f}\}}a_{k}^{\dagger }a_{k}$.

Obviously, there are only four eigenstates of $H_{0}^{XY}$ with zero
eigenvalues: the groundstate $|\psi _{g}^{XY}\rangle =\prod_{|k|\leq
k_{f}}a_{k}^{\dagger }|0\rangle $ and the eigenstate $|\psi
_{max}^{XY}\rangle =\prod_{|k|>k_{f}}a_{k}^{\dagger }|0\rangle $ of $%
H_{0}^{XY}$ with maximum eigenvalue. These two states indicate
that the two regions in $k$-space separated by fermi points are
fully filled respectively. If all the $k$-space is fully filled or
empty, one can get other two eigenstates $|\psi _{FM}\rangle $,
being the saturated ferromagnetic states with all spin up or down.
These four eigenstates are
also the eigenstates of any linear combinations of $H_{0}^{XY}$ and $%
H_{add}^{XY}$. Here we only consider the simplest case $%
H^{XY}=H_{0}^{XY}+H_{add}^{XY}$, that is
\begin{equation}
H^{XY}=2J\sum_{i}^{N}\cos ^{2}(\frac{\pi i}{N})(a_{i}^{\dagger
}a_{i+1}+h.c).
\end{equation}%
When $N/2$ is even, the $XY$ model is also equivalent to the
hardcore boson system. But we can not get the same analytical
result as that in the case of odd number of $N/2$. Although
numerical results for small $N$-site system show that the same
conclusion is also true, we can not give an exact proof at present
stage. Anyway, this sample implies that the factor $2\pi i/N$ is a
characteristic groundstate property for the spin ring system.

Finally we turn to the isotropic case, i.e. $\Delta =1$, in which
the Hamiltonians read as  $H_{0}=J\sum_{i}^{N}\mathbf{S}_{i}\cdot
\mathbf{S}_{i+1}$
and $H_{add}=J\sum_{i}^{N}\cos (2\pi i/N)\mathbf{S}_{i}\cdot \mathbf{S}%
_{i+1}.$ Correspondingly we have
\begin{equation}
H=H_{0}+H_{add}=2J\sum_{i}^{N}\cos ^{2}(\frac{\pi
i}{N})\mathbf{S}_{i}\cdot \mathbf{S}_{i+1}.
\end{equation}%
For $J<0$, it is easy to find that the ground states of $H_{0}$ and $H$ are
both saturated ferromagnetic with the same eigen energy $NJ/4$. The
saturated ferromagnetic state is also the eigenstate of $H_{add}$ with zero
eigenvalue.

For $J>0$, the saturated ferromagnetic state is also the common
eigenstate of $H_{0}$, $H_{add}$ and $H$. So part of the exact
conclusion obtained with $XY$ model can be extended to the
isotropic antiferromagnetic Heisenberg spin ring trivially. For
the ground state, we can not get similar analytical result as that
in $XY$ model. In order to investigate the relation between the
ground states of $H$ and $H_{0}$, exact diagonalization is
performed to compute the groundstate energies and the overlap of
the two groundstate wavefunctions. In Table 1, the numerical
results are listed for the systems of site number
$N=12,14,16,18,20,22$ and $24$. It shows that the overlap
approaches the unity, i.e. the ground states of these two
Hamiltonians are approximately identical.

\begin{center}
$%
\begin{tabular}{cccc}
\hline\hline
\ \ \ \ $N$ \ \ \  & $\ \ \ \Delta E\ \ \ \ $ & \ \ $Overlap$ \ \ \ \  &  \\
\hline
12 & 2.5$\times $10$^{-5}$ & 0.999988 &  \\
14 & 2.6$\times $10$^{-5}$ & 0.999983 &  \\
16 & 2.5$\times $10$^{-5}$ & 0.999978 &  \\
18 & 2.4$\times $10$^{-5}$ & 0.999974 &  \\
20 & 2.3$\times $10$^{-5}$ & 0.999971 &  \\
22 & 2.2$\times $10$^{-5}$ & 0.999967 &  \\
24 & 2.0$\times $10$^{-5}$ & 0.999964 &  \\ \hline
\end{tabular}%
$ \vspace*{1cm}

Table 1
\end{center}

\textit{Table 1. The differences of groundstate energies and the
overlaps of corresponding eigenfunctions of the Hamiltonian $H$
and $H_{0}$ for the N-site systems obtained by exact
diagonalization. It shows that the ground states of the two
Hamiltonians are approximately identical.} \newline
\newline

\vspace*{-4.0cm}

\begin{figure}[h]
\hspace{24pt}\includegraphics[width=9cm,height=12cm]{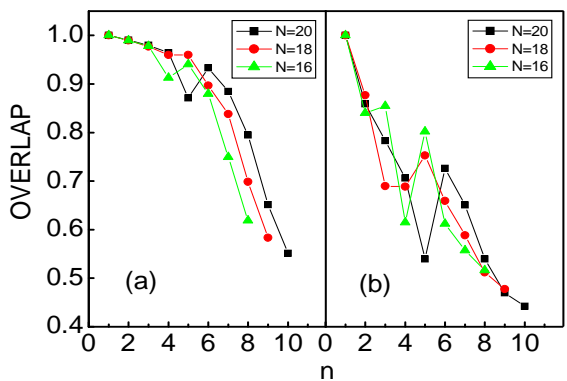} \vspace*{%
-3.5cm}
\caption{\textit{The overlaps between the groundstate wavefunctions of $%
H_{0}$ and $H_{add}$ with the coupling strength $J_i=J(1+0.5\cos(2n\protect\pi %
i/N))$(a) and $J_i=J(1+0.95\cos(2n\protect\pi i/N))$ (b) for
$N=20$, $18$ and $16$. It shows that only $2 \protect\pi i/N$
matches the ground state of uniform coupling system. }}
\end{figure}

The above results imply that the homogeneity of couplings is not
the unique optimal distribution for the translational invariant
groundstate
having maximal entanglement. Now we investigate the ground states of the $N$%
-site ring systems with the coupling strength $J_{i}=J+J'\cos
(2n\pi i/N))$, where $n=0,1,2,...,N-1$. Here $J' < J$ ensures $J_i
> 0$ in order to avoid the degeneracy of the ground
states~\cite{lieb}. Obviously, for $XY$ model
the conclusion for ground state obtained above is no longer available for $%
n\neq 0,1$. For isotropic Heisenberg model, a small size ring is
investigated by numerical method. In Fig.3, the overlap of the
corresponding eigenfunctions are plotted. It shows that $2\pi /N$
is the characteristic factor for the isotropic Heisenberg ring
system. On the other hand, from Fig.2 we can see that the
concurrence of spins at the two ends of a modulated chain should
be close to that of NN spins on a uniform ring.

\emph{Summary.} In summary, the correlation and entanglement of
the zero-spin eigenstates of the Heisenberg models are studied. We
find that the total NN correlation function of zero-spin
eigenstate reaches its local extremum when all coupling strengths
are identical. Applying this fact to $d$-D cubic AF Heisenberg
model, the groundstate concurrence, the measure of entanglement is
locally maximized at the same point. Numerical calculations are
employed to investigate a $N$-site quantum spin ring with
cosinusoidally modulated exchange couplings. It indicates that the
homogeneity of couplings is not the unique optimal distribution
for maximizing the groundstate entanglement and this modulation of
interactions can indeed result a longer rang entanglement.
Furthermore, it also implies that $2\pi /N$ is the characteristic
factor for the isotropic Heisenberg ringe system.

\textit{This work is supported by the NSFC and the Knowledge Innovation
Program (KIP) of the Chinese Academy of Sciences. It is also founded by the
National Fundamental Research Program of China with No. 2001CB309310. }

\end{document}